\documentclass{article}
\usepackage{spconf,amsmath,epsfig,subfig}
\usepackage{textpos}
\title{Aqua MODIS Electronic Crosstalk on SMWIR Bands 20 to 26}
\name{G. R. Keller$^{1}$, Z. Wang$^{1}$, A. Wu$^{1}$, X. Xiong$^{2}$}

\address{$^{1}$ Science Systems and Applications, Inc., Lanham, MD, 20706 USA.\\
$^{2}$ Sciences and Exploration Directorate, NASA Goddard Space Flight  Center, Greenbelt,  MD  20771  USA.}

\begin{document}

%\ninept
%

\maketitle

\begin{textblock*}{20cm}(-1cm,-7.0cm)
\footnotesize Copyright 2017 IEEE. Published in the IEEE 2017 International Geoscience \& Remote Sensing Symposium (IGARSS 2017), scheduled for July 23-28, 2017 in Fort Worth, Texas, USA. Personal use of this material is permitted. However, permission to reprint/republish this material for advertising or promotional purposes or for creating new collective works for resale or redistribution to servers or lists, or to reuse any copyrighted component of this work in other works, must be obtained from the IEEE. Contact: Manager, Copyrights and Permissions / IEEE Service Center / 445 Hoes Lane / P.O. Box 1331 / Piscataway, NJ 08855-1331, USA. Telephone: + Intl. 908-562-3966.
\end{textblock*}

\vspace{-0.5cm}
\begin{abstract}
Aqua MODIS Moon images obtained with bands 20 to 26 (3.66 -- 4.55 and 1.36 -- 1.39 $\mu$m) during scheduled lunar events 
show evidence of electronic crosstalk contamination of the response of detector 1.  
In this work, we determined the sending bands for each receiving band. 
We found that the contaminating signal originates, in all cases, from the detector 10 of the 
corresponding sending band and that the signals registered by the receiving and sending detectors 
are always read out in immediate sequence. We used the lunar images to derive the crosstalk coefficients, 
which were then applied in the correction of electronic crosstalk striping artifacts present in L1B images, 
successfully restoring product quality.

\end{abstract}
\section{Introduction}
\label{sec:intro}

The Moderate Resolution Imaging Spectroradiometers, or MODIS, on board NASA's low orbit Terra and Aqua
satellites were launched in 1999 and 2002, respectively. They are heritage Earth observing sensors and two of the most important instruments monitoring
ocean, land, and atmospheric processes and require very accurate and stable calibration. They acquire data in 16 mid- and long-wave infrared
thermal emissive bands (TEB) and 20 visible, near-, and short-wave infrared reflective solar bands (RSB). The
arrays of detectors comprising the MODIS' bands are placed on four separate focal plane assemblies (FPA) according to the
wavelength regime covered by the bands \cite{xiong2009overview,xiong2009aqua}.

The MODIS instruments have long been known to be subjected to electronic crosstalk (where signal transmitted in one channel creates an unsought effect in another channel) between bands in the same FPA \cite{xiong2007applications,sun2010terra,sun2014terra,rs8100806,truman2016}. 
However, a comprehensive effort in mapping the sending/receiving bands, deriving the crosstalk coefficients, and assessing the life-time impacts of the 
crosstalk contamination on the L1B product for bands 20 to 26, in the short- and mid-wave infrared (SMWIR) FPA had, so far, been lacking. 

In this paper, we work to fill this gap. We present evidence of electronic crosstalk contamination seen in Aqua MODIS Moon images 
from bands 20 to 26, obtained during scheduled lunar 
events, always affecting detector 1 (according to the so called `product order'). We identify the detectors and bands sending the 
contaminating signal, describe the mitigation strategy we adopted, and derive crosstalk coefficients for bands 20 to 26 from lunar events 
throughout the entire mission. We then link the crosstalk signatures on lunar images to striping artifacts present 
in L1B imagery, correct sample images and assess the impact of the contamination.

\section{Crosstalk Characterization from Lunar Observations}

Routine lunar observations, obtained for on-orbit radiometric calibration \cite{xiong2013modis}, can be used to identify signatures of electronic crosstalk contamination. 
These images are co-registered in all bands and the Moon appears in many consecutive scans. Because the consecutive scans overlap, it is then 
possible to construct images of the entire Moon from single detectors. Crosstalk contamination between bands is easy to spot on single detector 
images of the Moon, because both the main image and the so called crosstalk ghost are spatially restricted and can often be readily separated from one another. 
These ghost images of the Moon are much fainter than the main Moon image and are dislocated from it in both the along-scan and track directions, 
according to the distance between the receiving and sending bands in the FPA and on the detectors sending and receiving the contamination.  

\begin{figure*}[!t]
\centering
\includegraphics[width=\textwidth,height=2.9in]{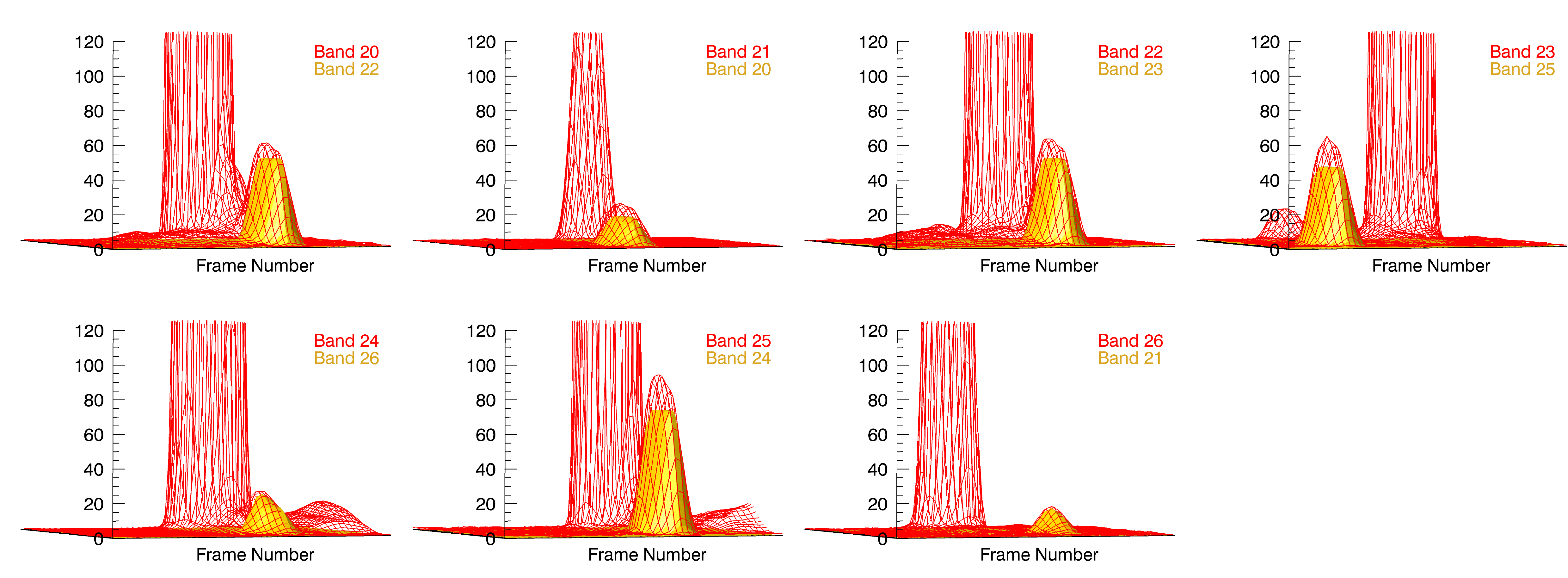}
\caption {{\small Crosstalk ghosts are present in detector 1 Moon images from bands 20-26 (red surfaces), obtained from a 2016 scheduled lunar event.
The images are displayed as surface plots and truncated in the $z$ axis for better visualization. The scaled Moon images from the detector 
10 of the corresponding sending bands
are overploted in yellow and precisely align with the crosstalk ghosts, after being dislocated in the along-scan direction of the 
appropriate number of pixels, corresponding
to the distances between sending and receiving bands in the FPA.}}
\label{fig_survey}
\end{figure*}

Fig. \ref{fig_survey} shows single detector Moon images from a 2016 Aqua MODIS scheduled event for bands 20-26, detector 1 (in red), as surface plots. 
All the images show the main lunar profile, truncated for clarity and at least one fainter ghost image. 
Although other ghosts are present, in this work, we will focus on electronic crosstalk ghosts that are apparent in
detector 1 Moon images and absent from images from other detectors and which, as we will see ahead, share a common mechanism. 
The surfaces in yellow are the scaled images from the sending band/detector. As it happens, the ghosts that only 
appear in detector 1 images are all caused by signal coming from detector 10 of the respective sending bands. 
Table \ref{tab_xtalk} lists the sending/receiving detector/band combinations we identified.

\begin{table} [!h]
\centering
\caption{\small{Electronic crosstalk sending/receiving band/detector combinations.}}
\begin{tabular}{ | c | c | c | c | c | c | c | c |}
\hline
Receiving Band     & 20 & 21 & 22 & 23 & 24 & 25 & 26 \\
and Detector &  1 &  1 &  1 &  1 &  1 &  1 &  1 \\
\hline
Sending Band       & 22 & 20 & 23 & 25 & 26 & 24 & 21 \\
and Detector   & 10 & 10 & 10 & 10 & 10 & 10 & 10 \\
\hline
\end{tabular}
\label{tab_xtalk}
\end{table}

\begin{figure}[!h]
\centering
\includegraphics[width=3.5in]{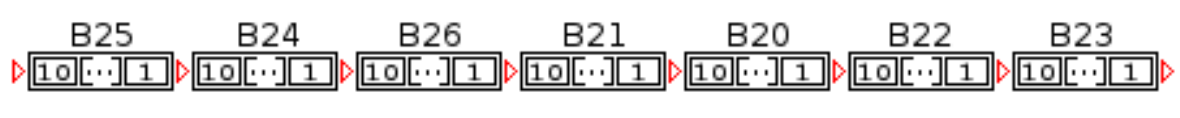}
\caption {{\small Read out order for bands in output 2 of the SMWIR FPA, starting from left to right. Band 25 is read again, after Band 23. Bands 5, 6, and 7, 
while in the same FPA, are not connected to the same electronic output.}}
\label{fig_sending}
\vspace*{-0.3cm}
\end{figure}

The fact that the ghosts in the images from the receiving bands align perfectly with the scaled Moon images from the sending bands after the predicted displacement in 
the along-scan direction is applied (according to the distance between bands in the FPA), together with the facts that the sending and receiving bands are all 
not only located in the same FPA, but connected to the same electronic output is strong evidence that the contamination is indeed caused by electronic crosstak. 
If we consider the band/detector order with which the signal of each of the bands connected to output 2 in the SMWIR FPA is read out, illustrated in Fig. \ref{fig_sending}, 
we find that the detector 1 of the receiving band is always read immediately before the detector 10 of the sending band. One possible interpretation is a failure 
in stopping the receiving band from being read out, by the time the signal from the next detector - from the sending band - starts being read. 
Bands 5, 6, and 7, while in the same FPA, are not connected to the same electronic output.

\section{Mitigation Strategy}

Following the literature \cite{sun2010terra,sun2014terra,truman2016}, we describe the contaminating signal according to the Equation \ref{eq_xt} 
\vspace*{-0.4cm}

\begin{equation}
dn_{r}(S,F) = dn_{r}^{*}(S,F) - c \times dn_{s}(S,F+\Delta F).
\label{eq_xt}
\end{equation}

Here, $dn$ denotes the background subtracted instrument response, the symbol * indicates the
contaminated signal, $c$ is the crosstalk coefficient and the subscripts $s$ and $r$ refer to the sending and receiving band/detector combination, respectively. The capital 
letters $S$ and $F$ denote scan and frame numbers, respectively, and $\Delta F$ denotes the displacement in the frame direction, which corresponds to the separation 
between the sending and receiving bands in the FPA.

\begin{figure}[!t]
\centering
\includegraphics[width=3.05in]{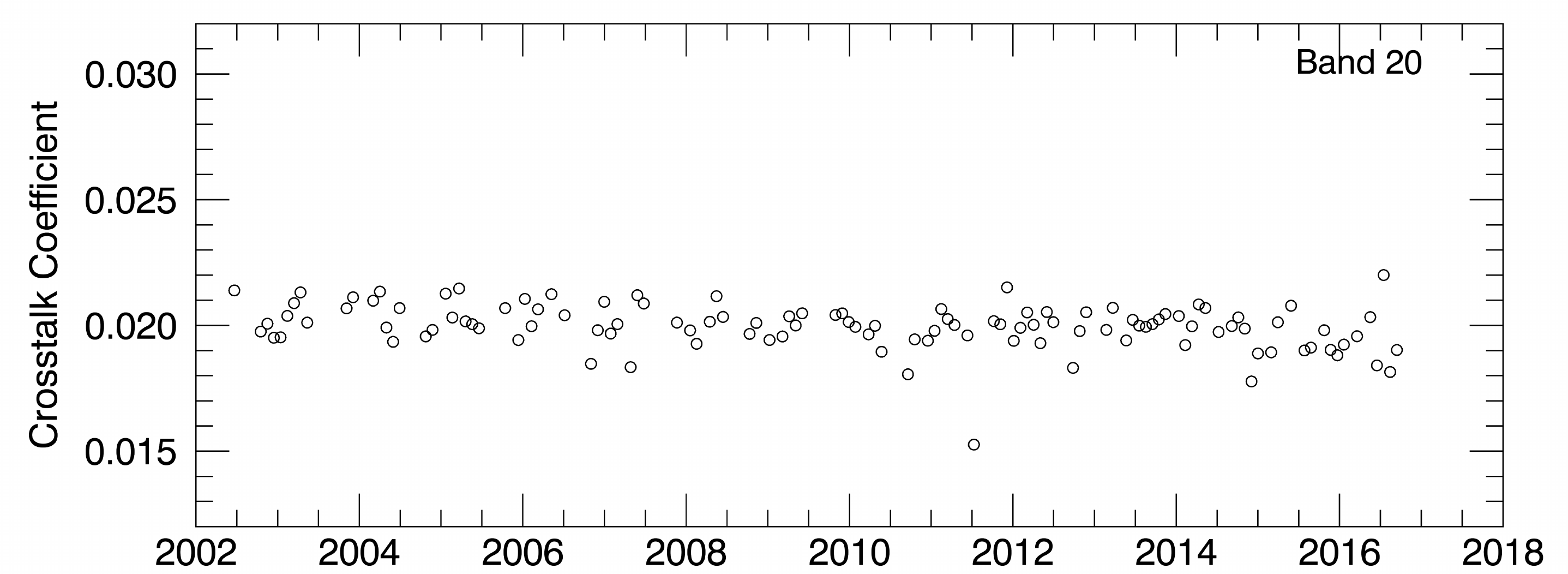}
\includegraphics[width=3.05in]{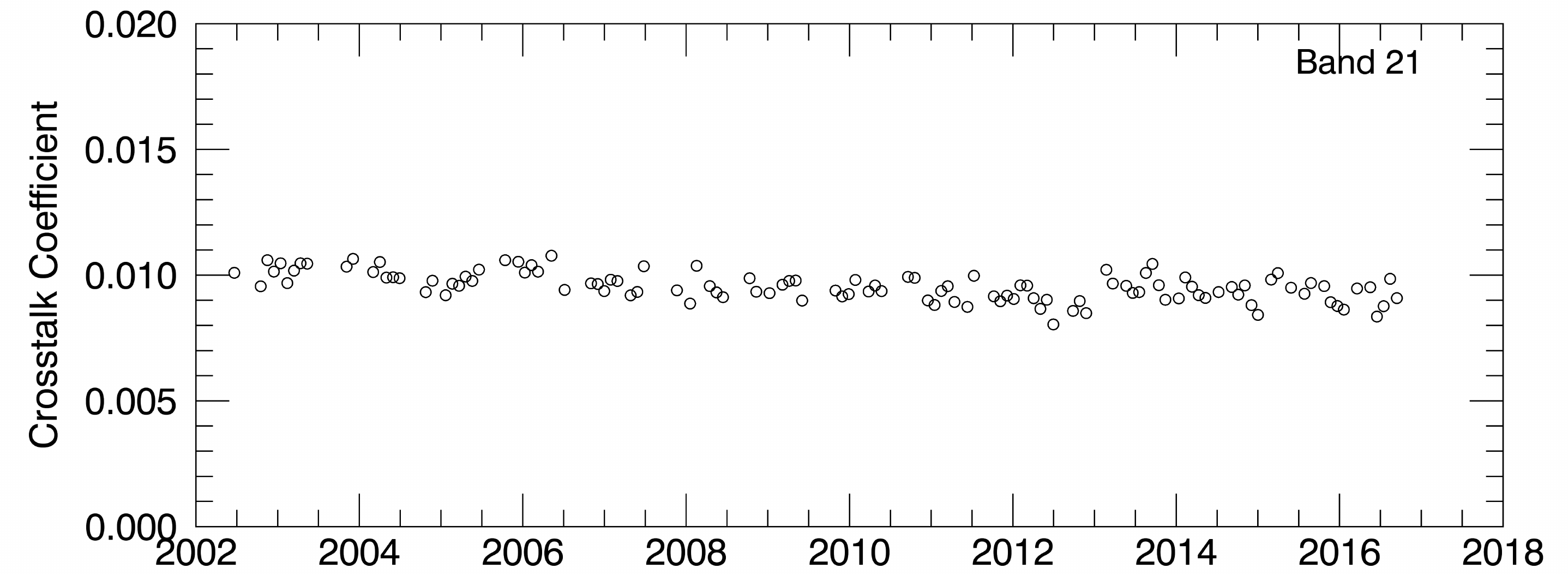}
\includegraphics[width=3.05in]{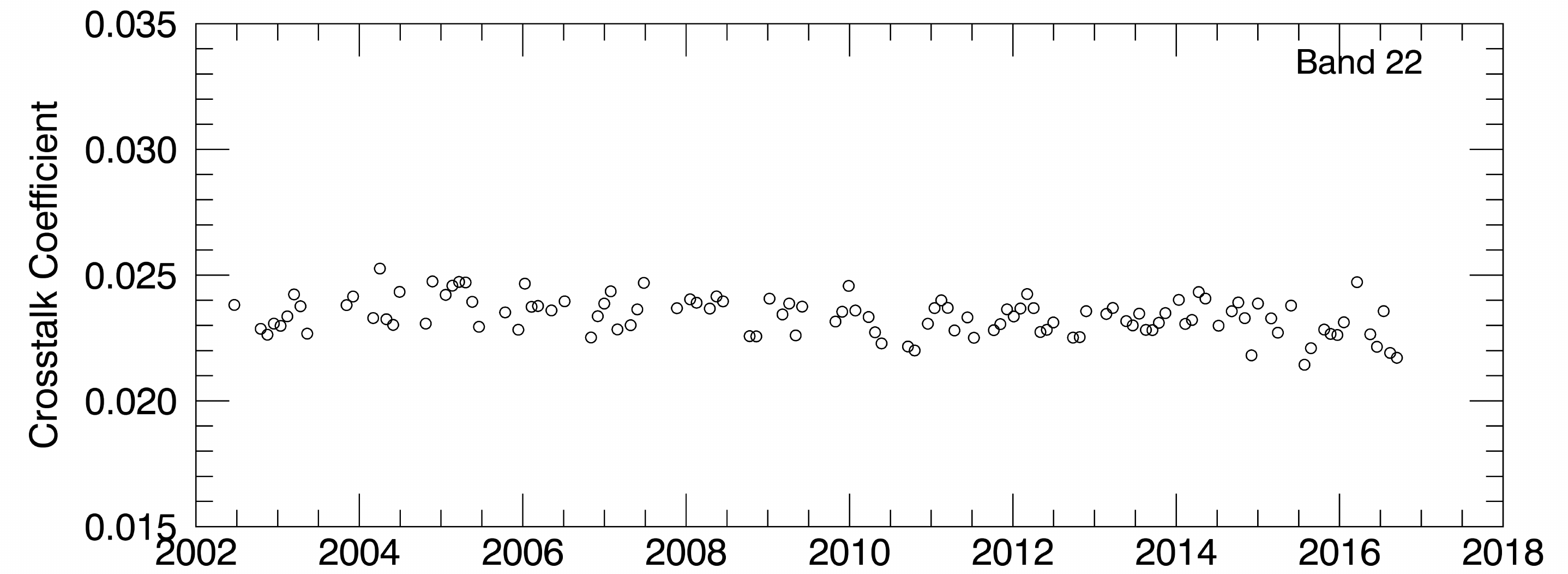}
\includegraphics[width=3.05in]{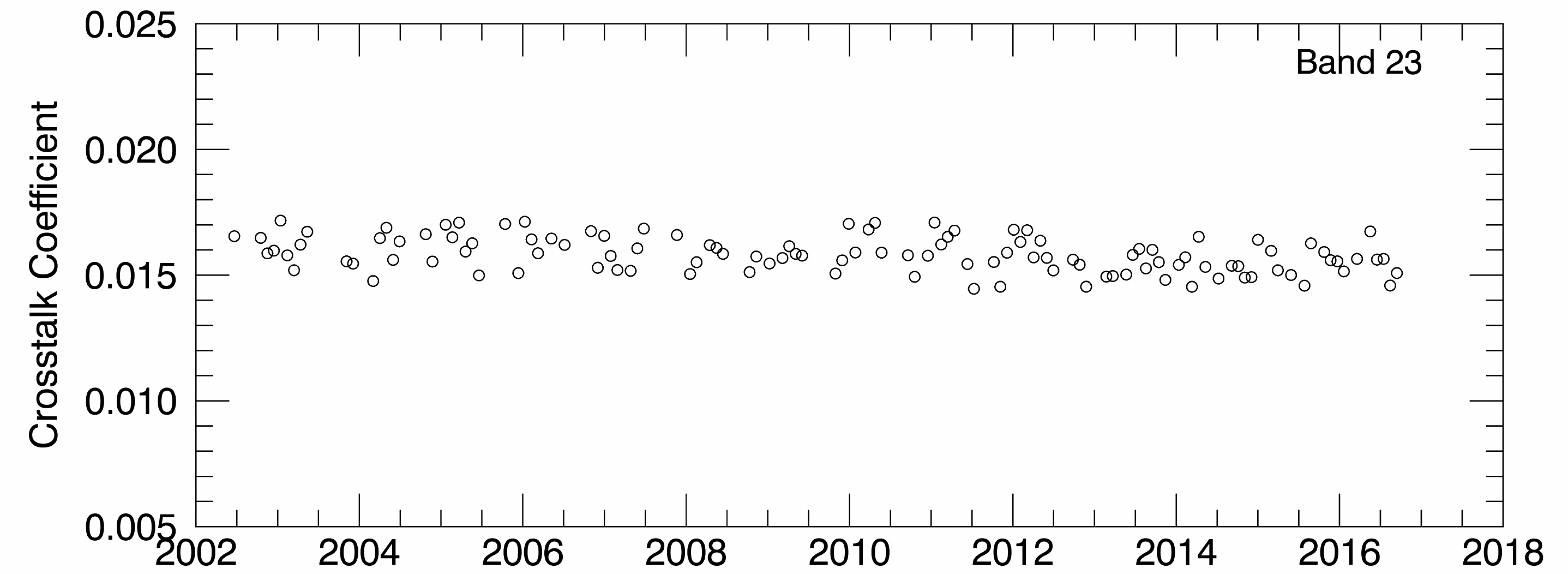}
\includegraphics[width=3.05in]{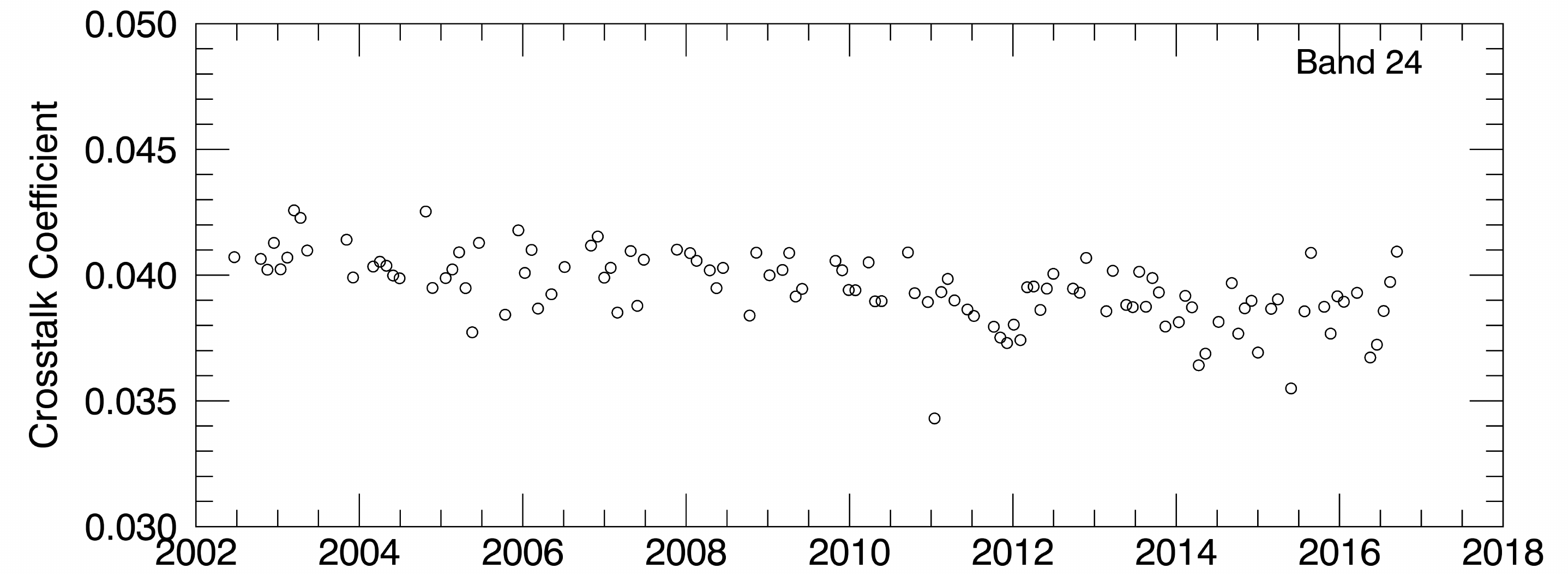}
\includegraphics[width=3.05in]{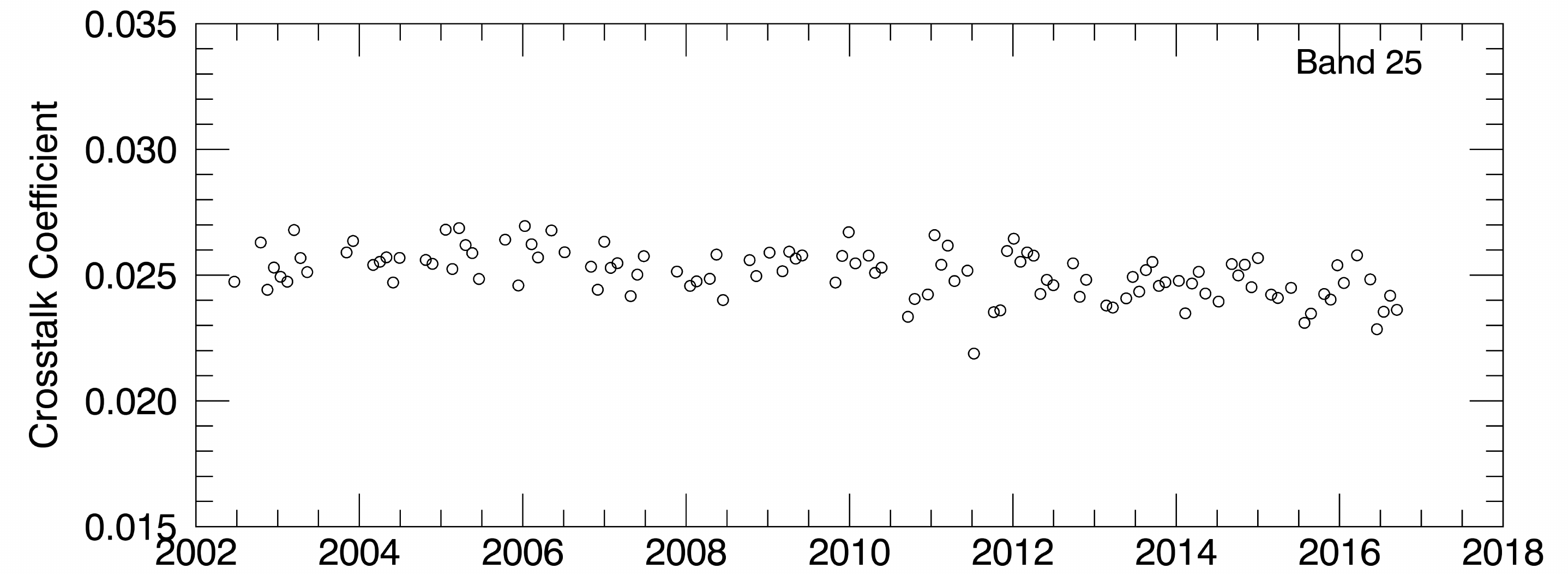}
\includegraphics[width=3.05in]{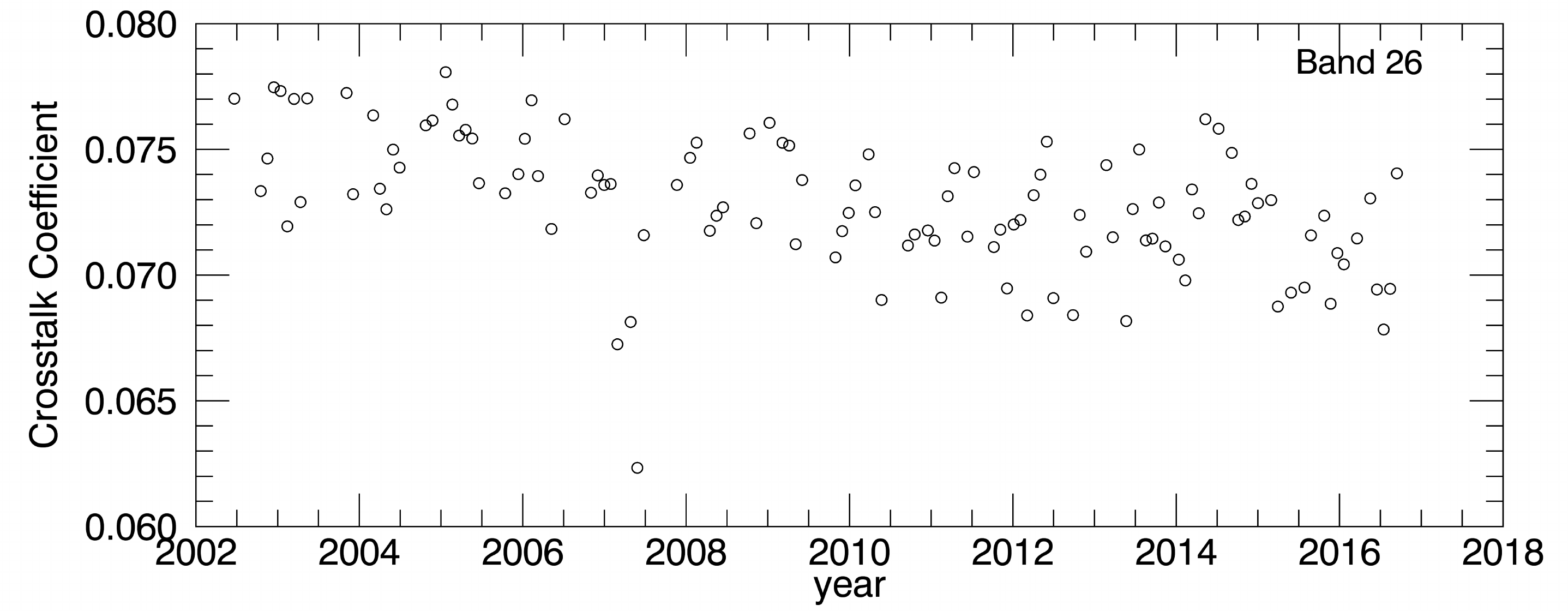}

\caption {{\small Crosstalk coefficients derived for scheduled lunar events over the entire mission, for the detector 1 of the receiving bands 20 -- 26 (corresponding sending bands in Table \ref{tab_xtalk}).}}
\label{fig_coeffs}
\end{figure}

The crosstalk coefficients are derived from the Moon images obtained during the scheduled lunar events throughout the mission and are shown in Fig. \ref{fig_coeffs}. 
For those sending bands where the Moon image saturates, we only considered the unsaturated pixels in the derivation of the crosstalk coefficient.    

\section{Impact on L1B}

\begin{figure}[!t]
\centering
\includegraphics[width=3.4in]{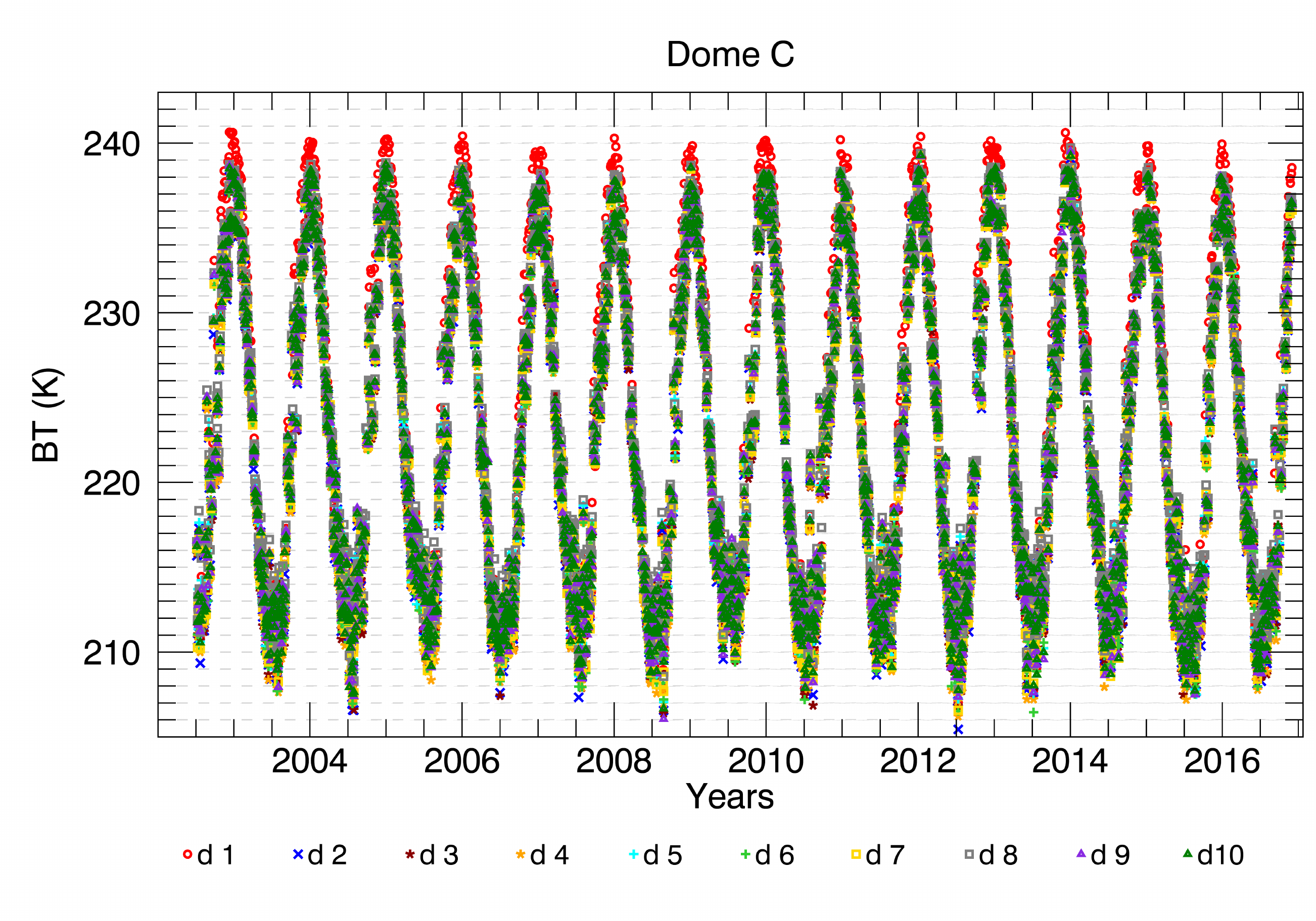}
\caption {{\small Dome C band 24 lifelong brightness temperature trend. Detector 1 from band 24 consistently reaches temperatures 2 K higher than the other detectors, around summertime.}
}
\label{fig_domeC}
\vspace*{-0.3cm}
\end{figure}

\begin{figure}[!t]
\centering
\includegraphics[width=3.3in,height=1.5in]{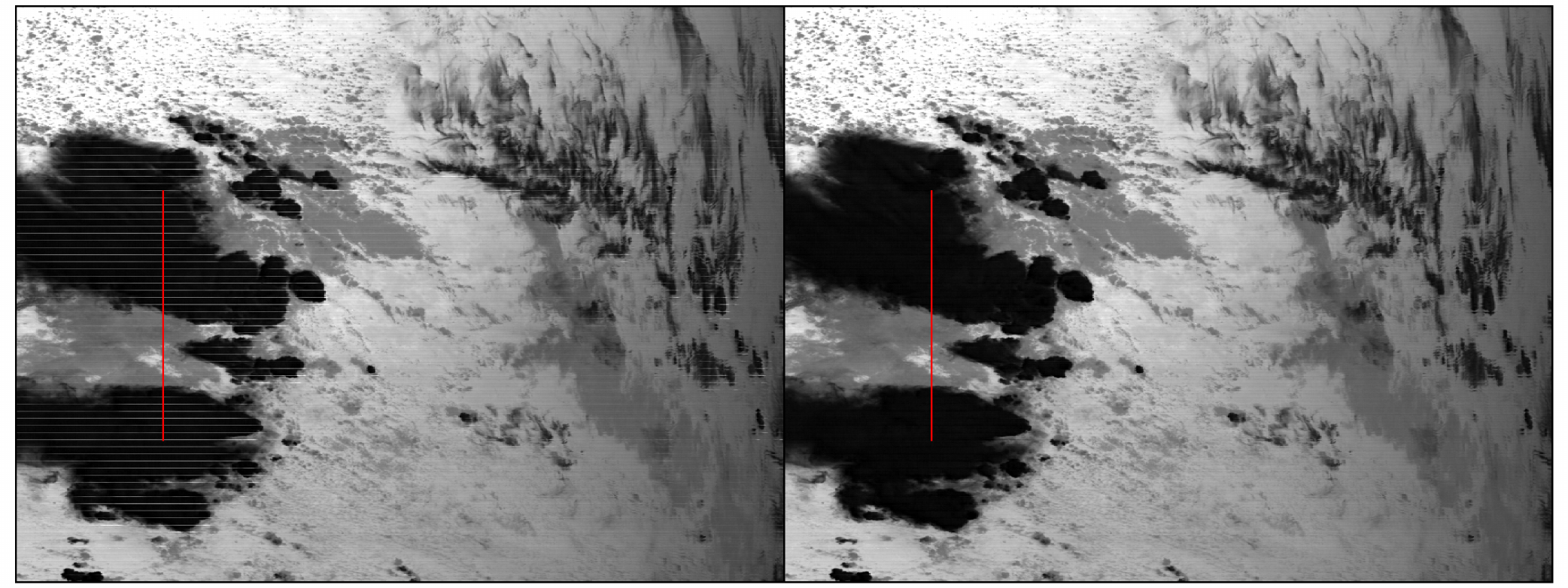}
\includegraphics[width=3.3in,height=1.5in]{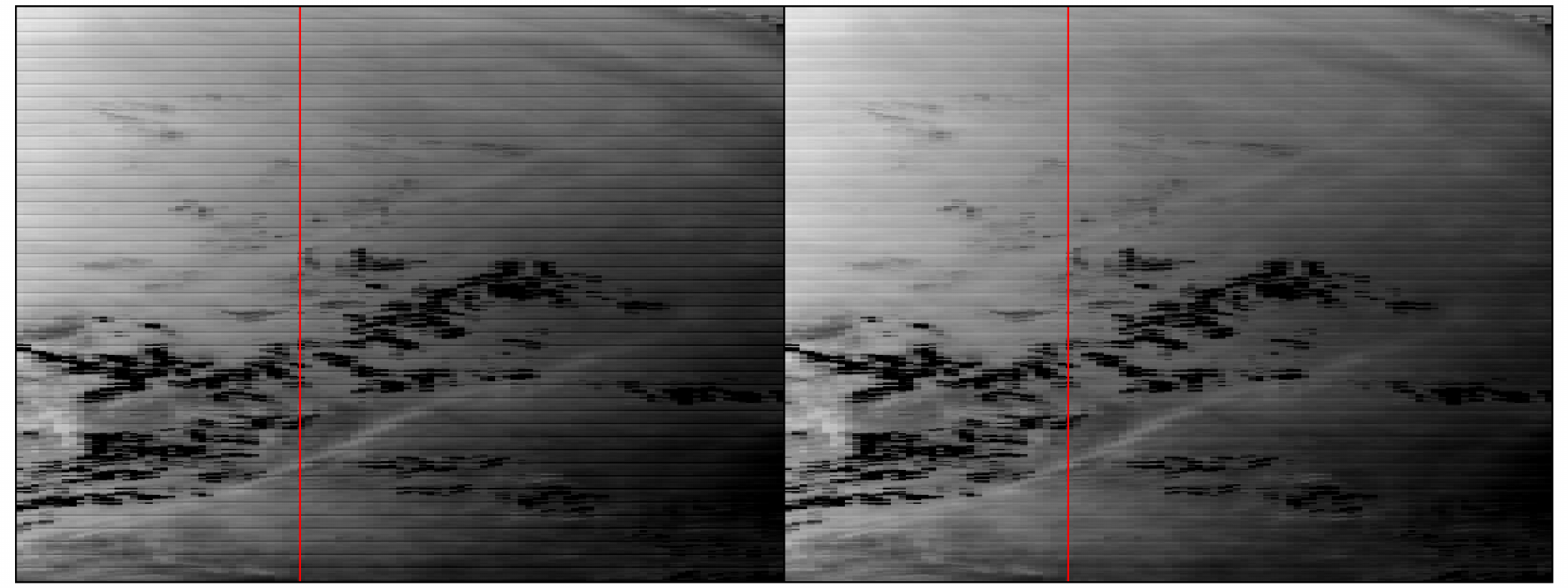}

\caption {{\small Sample L1B images from Aqua MODIS Bands 24 (top) and 23 (bottom), before (left) and after (right) crosstalk correction. 
The red lines indicate the regions used to derive the intensity profiles shown in Fig. \ref{fig_profile}. Fainter striping artifacts in other detectors 
are present, but were not addressed here.}
}
\label{fig_strips}
\end{figure}

The crosstalk contamination can potentially affect the L1B imagery as is evident in the lifelong brightness temperature trend of the Dome
Concordia region (Fig. \ref{fig_domeC}), which shows the detector 1 of band 24 consistently reaching temperatures 2 K higher than all the other detectors during summertime and in the band 24 L1B 
image shown in the top panel of Fig. \ref{fig_strips}, which shows distinct striping. Although less prominent, striping artifacts linked to the crosstalk are also present in other bands, 
for example, bands 23 (also shown in Fig. \ref{fig_strips}) and 25. Details on the electronic crosstalk contamination of Band 24 images can be found on \cite{keller2017aqua}.

\begin{figure}[!t]
\centering
\includegraphics[width=3.0in]{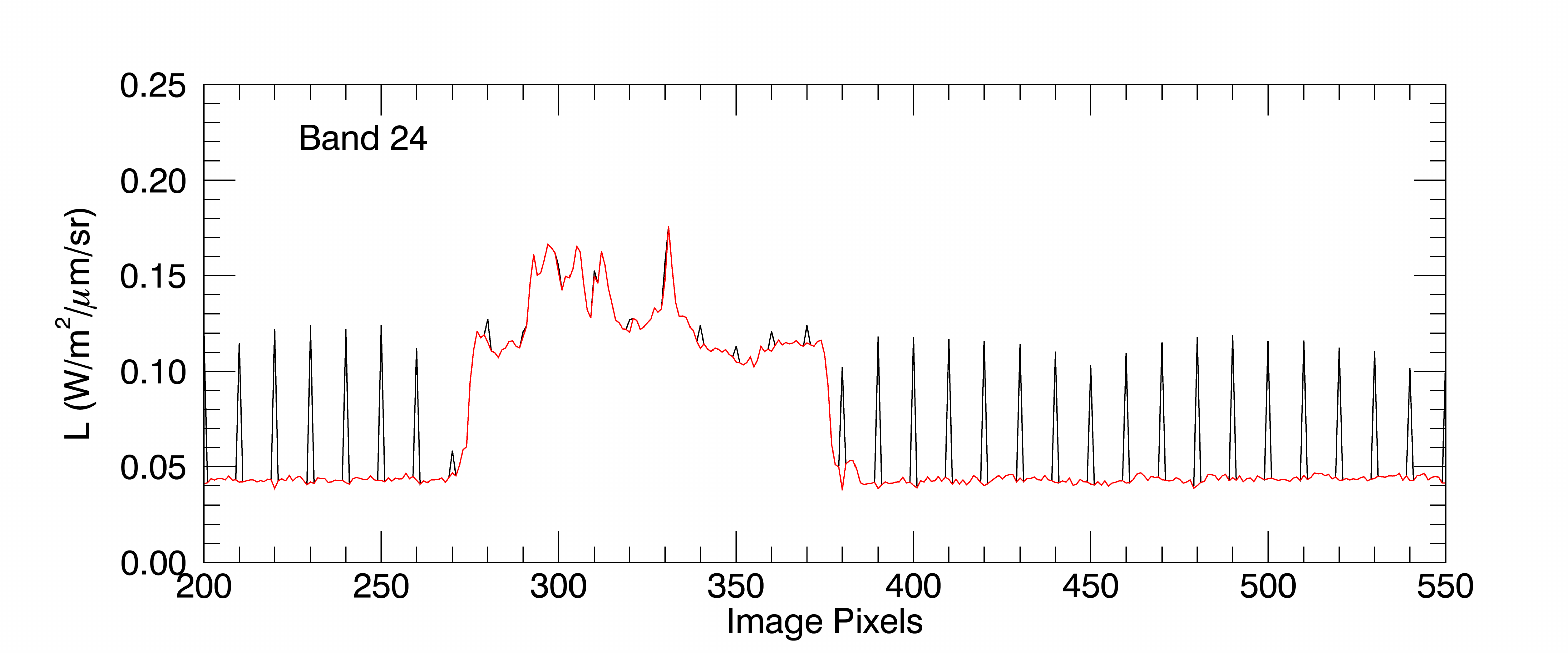}
\includegraphics[width=3.0in]{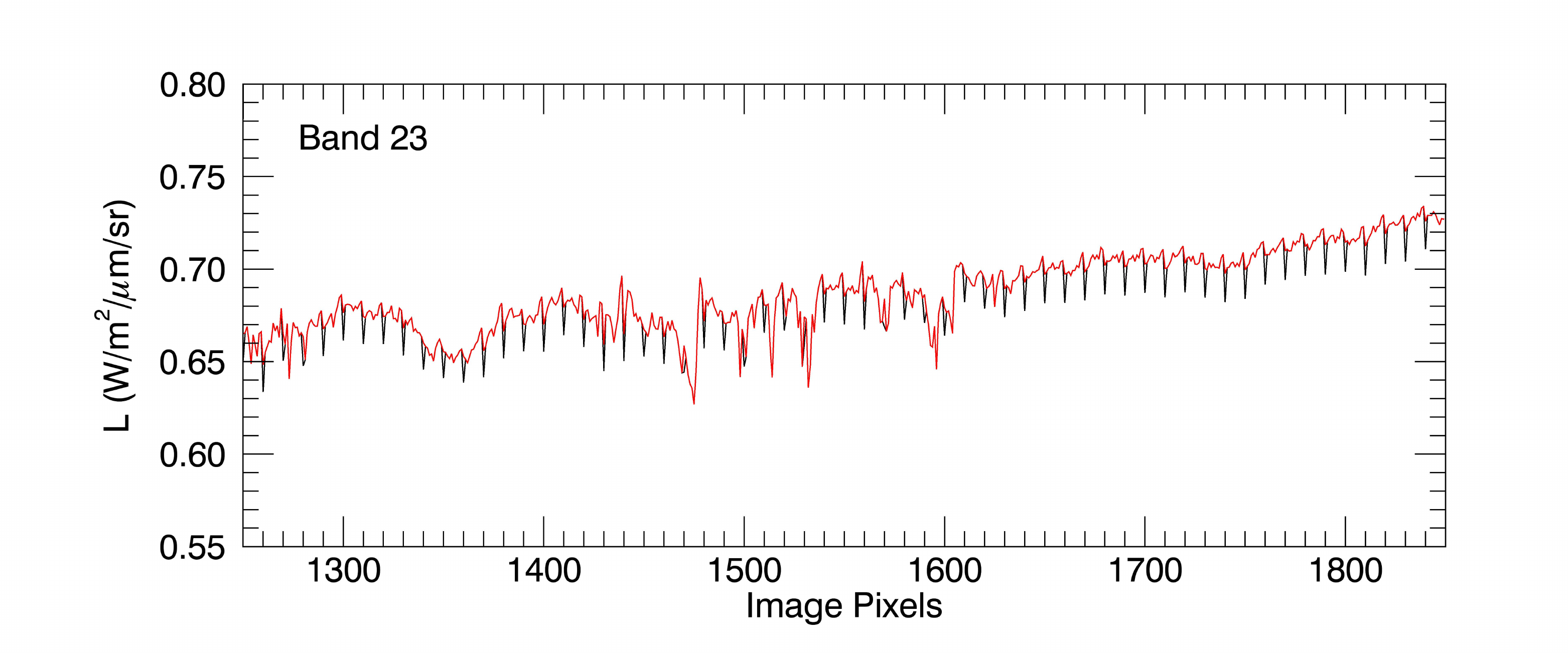}
\includegraphics[width=3.0in]{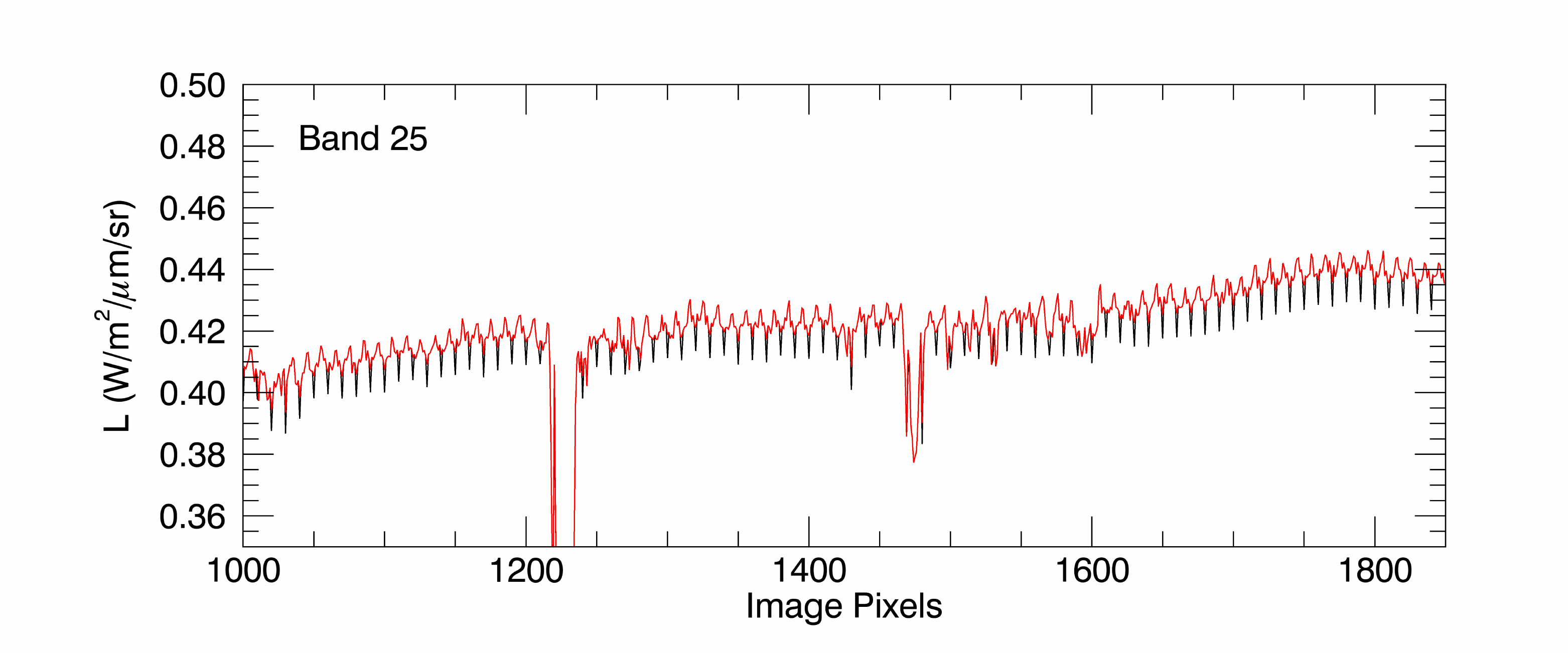}

\caption {{\small Intensity profiles extracted from the regions marked with red lines on the images in Fig. \ref{fig_strips} 
before (black) and after (red) correction. The image corresponding to the profile of band 25 was not shown. However, the profile was taken from 
the same region as the one marked on the image from Band 23.}}
\label{fig_profile}
\end{figure}

Once the crosstalk coefficients are derived, they can be applied in the correction of L1B imagery, as illustrated in Figs. \ref{fig_strips} and \ref{fig_profile}, which 
demonstrate the ability of the mitigation strategy in restoring product accuracy and image quality.

\section{Conclusion}

In this work, we provided evidence of electronic crosstalk contamination of the detector 1 signal in bands 20 to 26 using images of the Moon.
We identified the corresponding sending bands and determined that in all the cases studied the signal originates from detector 10. We
found the signal of the receiving and sending bands/detectors to always be read sequentially. One possible interpretation for this would be a failure
to stop the receiving band from reading out signal timely.

The mitigation strategy adopted in this work consists of modeling the contaminating signal as being proportional to the signal from the sending detector and
deriving the crosstalk coefficients from lunar images. We derived crosstalk coefficients for bands 20 to 26, for the entire mission.

We linked striping artifacts observed in L1B imagery to the crosstalk signatures in lunar images and corrected
sample L1B images to illustrate the ability of our approach in mitigating crosstalk artifacts, and in improving product accuracy.

A comprehensive study of the impacts of crosstalk contamination on the L1B product for bands 20 to 26 will be presented in a future work.

\section{Acknowledgements}
The authors thank other members of MCST.
%\textbf{Acknowledgement:} The authors thank other members of MCST.

\bibliographystyle{IEEEbib}
\bibliography{IEEEabrv,gk_refs}

\end{document}